\RequirePackage{fix-cm}
\documentclass{svjour3}                     % onecolumn (standard format)
\smartqed  % flush right qed marks, e.g. at end of proof
\usepackage{graphicx}
\usepackage{latexsym}
\newsavebox{\astrutbox}
\sbox{\astrutbox}{\rule[-5pt]{0pt}{20pt}}

\def\der#1#2{{\partial #1\over \partial #2}}

\def\be{\begin{equation}}
\def\ee{\end{equation}}
\def\bea{\begin{eqnarray}}
\def\eea{\end{eqnarray}}
\def\bse{\begin{subequations}}
\def\ese{\end{subequations}}
\def\bsea{\begin{subeqnarray}}
\def\esea{\end{subeqnarray}}
\def\({\left (}
\def\){\right )}
\def\[{\left [}
\def\]{\right ]}
\def\<{\left <}
\def\>{\right >}

\begin{document}

\title{Quantifying some properties of the stochastic quantum force}

\titlerunning{}

\author{Roumen Tsekov and Eyal Heifetz}


\institute{Roumen Tsekov \at Department of Physical Chemistry, University of Sofia, 1164 Sofia, Bulgaria\\
{\email Tsekov@chem.uni-sofia.bg}\ \and
Eyal Heifetz \at Department of Geosciences, Tel-Aviv University, Tel-Aviv, Israel\\
  {\email eyalh@post.tau.ac.il} }


\date{\today}

\maketitle
\begin{abstract}

We consider for clarity the simple case of the one dimensional non-relativistic Schr\"{o}dinger equation and regard it as an ensemble mean representation of the stochastic motion of a single particle in a vacuum, subject to an undefined stochastic quantum force. By analyzing the Bohm potential it is found that the imaginary part of the quantum momentum is the root mean square fluctuation of the particle around its mean velocity, where the latter is the real part of the quantum momentum. The local mean of the quantum force is found to be proportional to the third spatial derivative of the probability density function, and its associated pressure to the second spatial derivative. The latter is decomposed from the single particle diluted gas pressure, and this pressure partition allows the interpretation of the quantum Bohm potential as the energy required to put a particle in a bath of fluctuating vacuum at constant entropy and volume.

The stochastic force expectation value is zero and is uncorrelated with the particle location, thus does not perform work on average. Nonetheless it is anti-correlated with volume and this anti-correlation leads to a new type of an Heisenberg like relation. We imply the dynamic Gaussian solution to the Schr\"{o}dinger equation as a simple example for exploring the mean properties of this quantum force. Still an interesting remained open task is the identification of the stochastic law that leads to these obtained mean properties.

\end{abstract}
%
%
%
%\section{Introduction}
%%\label{intro}

\section{Introduction}

Since the seminal work of Nelson \cite{Nelson66}, there is a continuous attempt to show that the Schr\"{o}dinger equation (hereafter SE) describes the ensemble mean dynamics of particles which experience stochastic motion due to the interaction with the vacuum. The recent book on ``The Emerging Quantum'' \cite{Emerg2015} gathers the work along this line throughout the years and the reader is kindly referred to this book and to its extensive reference list. The work presented here is an additional step toward the characterization of the mean properties of this stochastic quantum force.

In the heart of the analysis is the assumption that the motion of a particle in a vacuum satisfies the Newton second law
\be
m\ddot{\bf X} = -\nabla{U} + {\bf F} ,
\ee
where ${\bf X}(t)$ is the particle location at time $t$, $U$ is an external potential and ${\bf F}$ is an undefined stochastic force acting on the particle. 
Since SE predicts the probability density function (PDF) $\rho({\bf x},t)$, obtained from many repetitions of an identical experiment of a particle in a vacuum, we wish to relate SE to (1) by looking at the $n^{th}$ realization of the particle location ${\bf X}_n$ and expect that the PDF
\be
\rho({\bf x},t) = \lim_{N \rightarrow \infty}{1\over N}\sum_{n=1}^{N}\delta({\bf x}-{\bf X}_n),
\ee
(where $\delta$ denotes the Dirac delta function) obtained from the solution of (1), will agree with the PDF obtained from the solution of SE
\be
i\hbar\der{\Psi}{t} = \hat{H}\Psi = \({\hat{p}^2 \over 2m} +U\)\Psi =
\(-{\hbar^2\over 2m}\nabla^2 +U\)\Psi,
\ee
with the wave function $\Psi$ written in the polar form
\be
\Psi({\bf x},t) = \sqrt{\rho}({\bf x},t)e^{iS({\bf x},t)/\hbar}
\ee
and ${\hat p} = -i\hbar \nabla$.

The rest of this short paper is organized as follows. In Section 2 we compare the Madelung equations, derived from SE, with the general fluid dynamics Euler-like equations, resulting from the ensemble mean of (1). This comparison (in 1D for simplicity) enables us to obtain the mean expressions for the stochastic quantum motion of (1) and especially for the force  ${\bf F}$. In Section 3 we analyze the associated properties of the stochastic quantum force, where in Section 4 we imply the dynamic Gaussian solution of the 1D SE to obtain a simple example of this mean stochastic dynamics. We close in Section 5 by discussing the results. 

%Since the work of Nelson \cite{ }, there is an attempt to show that the Schr\"{o}dinger equation describes the ensemble mean dynamics of particles experiencing stochastic motion (add refs \cite{ , , ,}...). This approach is fundamentally different from the far more accepted wave-particle duality one, however it is in a sense easier to digest -- if for some reason a particle in a vacuum is subject to some stochastic forcing, a probability density function (pdf) equation is a natural way to describe its mean motion. Obviously, in order to prove that this is indeed the case, one has to first identify the governing law of this stochastic dynamics, and then to explain why the wave-particle duality description leads to the same equation. 
%
%To the best of our knowledge neither of these two tasks have been accomplished in a satisfied way. Here, we suggest a step in the direction of the first task.
%We consider for clarity the simplest case of the 1D non-relativistic Schr\"{o}dinger equation and present it as an ensemble mean of undefined stochastic motions of identical set of particles in a vacuum. Then the two components of the Bohm potential allow us to determine the root mean square velocity and the mean stochastic force acting on the particle.
%
%Still a remaining open question is the 
   %

\section{Ensemble mean quantum dynamics}

The Madelung \cite{Mad} equations are obtained when taking the real and the imaginary parts of SE. They become respectively continuity and momentum like equations, where in 1D they take the form of
\be
\der{\rho}{t} =-\der{}{x}(\rho u),
\ee 
\be
m{Du\over Dt} = -\der{}{x}(Q+U),
\ee
where 
\be
{u} = {1\over m}\der{S}{x} = {1\over m} \Re{\({\hat{p}\,\Psi \over \Psi}\)}, 
\ee 
\be
{Q}  = - {\hbar^2 \over 2m\sqrt{\rho}}{\der{^2\sqrt{\rho}}{x^2} }
\ee
is the Bohm potential \cite{Bohm} and ${D\over Dt} \equiv \der{}{t}+ u\der{}{x}$ is the material derivative in 1D.
We note as well that the energy expression
\be 
E = \int_{-\infty}^{\infty}\(-{\hbar^2\over 2m}\Psi^{*}\der{^2 \Psi}{x^2}+  U|\Psi|^2\) dx = 
\int_{-\infty}^{\infty} \rho \[ {m\over 2}(u^2 +v^2) +U\]dx, 
\ee
combines the mean kinetic energy $K={mu^2\over 2}$ as well as the internal energy $I = {mv^2\over 2}$. The latter is proportional to the Fisher information \cite{HC}
with ``diffusive velocity'' \cite{Entropy} defined as
\be
v = -{\hbar \over 2m}\der{}{x}\ln{\rho} = {1\over m} \Im{\({\hat{p}\,\Psi \over \Psi}\)}.
\ee

The plan now is to obtain the ensemble mean equations resulted from (1), for the 1D case, and then compare them with the Madelung equations. 
To be more explicit, we consider a single particle moving stochastically in a vacuum obeying to (1). We mentally repeat this experiment ${N \rightarrow \infty}$ times where the experiments are independent from each other. The 1D version of (2) is then the PDF of the particle position so that  
\be
\int_{- \infty}^{\infty}\rho(x,t)dx = 1.
\ee
Defining for convenience the global and local averaging operators as
\be
\<f\> \equiv \lim_{N \rightarrow \infty}{1\over N}\sum_{n=1}^{N} f_n\, ; \hspace{0.2cm}
{\overline f} \equiv \[\lim_{N \rightarrow \infty}{1\over N}\sum_{n=1}^{N} f_n \delta({x}-{X}_n)\] /
\[\lim_{N \rightarrow \infty}{1\over N}\sum_{n=1}^{N} \delta({x}-{X}_n)\] 
\ee
then
\be
\rho(x,t) = \< \delta(x-X) \>\, ; \hspace{0.2cm} {\overline f}(x,t) = {\<f\delta(x-X) \> / \rho} \hspace{0.1cm} \Rightarrow \hspace{0.1cm}
\<f\> =  \int_{-\infty}^{\infty} \rho  {\overline f} dx.
\ee
We wish to obtain a continuity like equation from the stochastic dynamics. Taking the time derivative of $\rho$ we get 
\be
\der{\rho}{t} = \<\der{}{t} \delta(x-X) \> = \<\dot{X} \der{}{X} \delta(x-X) \> 
= - \der{}{x} \<\dot{X}\delta(x-X) \>.
\ee
Using the local averaging operator of (13), (14) becomes the continuity equation (5) with the ensemble mean velocity 
\be
u(x,t) = \overline{\dot X},
\ee
which can be  equated to the de Broglie guiding velocity in (7). 

Next we derive an ensemble mean momentum like equation. Toward this end we take the time derivative of the momentum flux
\be
\der{}{t}(\rho u) = \<\ddot{X}\delta(x-X)\> - \der{}{x}\<\dot{X}^2\delta(x-X)\>
\ee
but when using the continuity equation (5) we also get 
\be
\der{}{t}(\rho u) = \rho\(\der{u}{t}+ u\der{u}{x} \) - \der{}{x}(\rho u^2) \equiv \rho{Du\over Dt} - \der{}{x}(\rho u^2).
\ee
Equating between the RHS of (16) and (17) we obtain
\be
{Du\over Dt} = \overline{\ddot{X}} - {1\over \rho}\der{}{x}\[\rho \overline{(\dot{X}-u)^2}\].
\ee
Equating between (6) and (18) together with the partition of the Bohm potential gradient  to
\be
-{1\over m}\der{Q}{x} = \({\hbar \over 2m}\)^2  {1 \over \rho} \der{^3\rho}{x^3} - {1\over \rho}\der{}{x}\(\rho v^2\),
\ee
suggest that
\be 
 \overline{(\dot{X}-u)^2} = v^2,
\ee
and
\be
m\overline{\ddot{X}} = -\der{U}{x} + {m}\({\hbar \over 2m}\)^2  {1 \over \rho} \der{^3\rho}{x^3}.
\ee
Equating (21) with the ensemble mean of the 1D version of (1) we obtain the expression for the local mean srochastic quantum force 
\be 
\overline{F} = {m}\({\hbar \over 2m}\)^2  {1 \over \rho} \der{^3\rho}{x^3}.
\ee
For consistency we note as well that using (13) and (20) it is straightforward to see that the global energy of the particle satisfies
\be 
\<m\({\dot{X}^2\over 2} + U\)\> = \int_{-\infty}^{\infty} \rho m\({\overline{\dot{X}^2}\over 2}+ U\) dx = 
\int_{-\infty}^{\infty} \rho m\[{(u^2 +v^2)\over 2}+U\]dx = \<K+I+U\>, 
\ee
in agreement with the RHS of (9).

\section{Properties of the stochastic quantum force}

We derived the ensemble mean expressions for the stochastic motion of a single particle in a vacuum when the experiment is independently repeated an $N$ large number of times. This statistical properties are equivalent to the ones of $N$ co-existing, non interacting, identical particles moving stochastically. In this context, one can write (18) as an Euler-like fluid momentum equation
\be 
m{Du\over Dt} =  - {1\over \rho}\der{p}{x} -\der{U}{x},
\ee
where (19-21) indicate that the pressure consists of the two following terms
\be 
{p\over \rho} = -{1\over m}\({\hbar \over 2}\)^2\der{^2 \ln \rho}{x^2} = m v^2 -{1\over \rho m}\({\hbar \over 2}\)^2 \der{^2{\rho}}{x^2} 
\equiv {1\over \rho}\(p_g +p_v\).
\ee
In the classic kinetic theory of gases $m\overline{\ddot{X}} = -\der{U}{x}$, i.e. the local mean stochastic force is zero, hence $p = p_g$. 
Here $p_g$ can be regarded as the pressure of an extremely diluted ideal gas, consisting of a single particle moving stochastically. For 1D ideal gas\, 
$p_g/\rho = m v^2 = k_B T = 2I$, where $T$ is the ideal gas temperature and $k_b$ is the Boltzmann constant. The translation of SE to the fluid like equation of 
(24) suggests an external non positive definite pressure term
\be 
p_v = -{1\over m}\({\hbar \over 2}\)^2\der{^2{\rho}}{x^2}, 
\ee
acting on the particle due to the vacuum fluctuations. Furthermore, from (8), (10) and (25) we obtain that
\be 
Q = I +{p_v \over \rho},
\ee
suggesting that $Q$ can be regarded as the part of the enthalpy associated with the vacuum fluctuations. Since both the temperature and the entropy in vacuum are zero, this enthalpy equals to the Gibbs energy which is equal as well to the chemical potential. Therefore, the quantum Bohm potential can be interpreted as the energy required to put a particle in a bath of fluctuating vacuum at constant entropy and volume.

It is straightforward to verify that the global mean of the stochastic quantum force is zero ($\<F\>=0$). 
%Moreover, equations (26) suggests a definition to the energy attributed from the vacuum pressure 
%\be
%{E}_v \equiv {p_v\over 2 \rho} = {{\hat p}^2 \rho \over 8m \rho} 
%\ee
%(please note that ${\hat p} = -i\hbar \der{}{x}$ is the momentum operator in 1D, not to be confused with the pressure $p$), yielding $\<{E}_v \>=0$.
Information on its higher moments can be obtained when applying (13) on (22) to obtain
\be
\<F\delta(x-X)\> = {m}\({\hbar \over 2m}\)^2 \der{^3\rho}{x^3}.
\ee
Implementing then the Fourier transform 
\be
{\tilde f}(q) = \int_{- \infty}^{\infty} f(x) e^{-iqx}dx,
\ee
so that
\be
{\tilde \rho} = \<e^{-iqX}\> = \sum_{k=0}^{\infty}{(-iq)^k \over k!}\<X^k\>,  
\ee   
the Fourier transform of (28) gives
\be
\sum_{k=0}^{\infty}{(-iq)^k \over k!}\<X^k F\> = -{1\over m}\({\hbar \over 2}\)^2 \sum_{n=0}^{\infty}{(-iq)^{n+3} \over n!}\<X^n\>,
\ee
yielding $\<{F} \> = \<X {F} \> = \<X^2 {F} \> =0$. The second term shows that the mean work performed by the stochastic quantum force is zero, hence has zero contribution to the energy integral of (23).  
Comparing in (31) the $k=3$ term at the LHS with the $n=0$ one at the RHS gives 
\be
\<X^3 {F} \> = -{3\hbar^2\over 2m}.
\ee
This equality is in a sense a new type of an Heisenberg-like relation. For Gaussian processes (Section 4) it can related to the Heisenberg uncertainty inequality.  
Note that since by definition $\<X\>=0\,$ for stochastic motion this implies as well that $\<X^4 {F} \>  = 0$. For completeness, the general relation for higher moments is given by
\be
\<X^{k+3} {F} \> = -{1\over m}\({\hbar \over 2}\)^2 (k+1)(k+2)(k+3)\<X^k\>. 
\ee

\section{Dynamic Gaussian solution}

To obtain some familiarity with these results we consider the dynamic Gaussian solution of the 1D SE \cite{Entropy,Tsekov12} for a free particle
\be 
\rho(x,t) =  {1\over \sigma(t) \sqrt{2\pi}}e^{-{x^2 \over 2\sigma^2(t)}}.
\ee
Substituting (34) in (5) and (6) yields
\be
u(x,t) = x\der{\ln\sigma}{t}, 
\ee
\be
\sigma\der{^2\sigma}{t^2} = \({\hbar \over 2m\sigma}\)^2 \hspace{0.25cm} \Longrightarrow  \hspace{0.25cm}
\sigma^2 = \sigma_0^2 + \({\hbar t\over 2m\sigma_0}\)^2.
\ee 
Substitute (36) back in (35) and in (10) we then obtain
\be
u = \({\hbar \over 2m\sigma_0}\)^2{xt \over \sigma^2} =  {xt \over \({m\sigma_0^2\over \hbar}\)^2 +t^2}\, ; \hspace{0.25cm}
v = {\hbar x \over 2m \sigma^2} = \({2m \sigma_0^2\over \hbar}\){u\over t}.
\ee
Hence while at short times $u \sim xt$ and $v \sim x$, at large times $u \sim x/t$ and $v \sim x/t^2$.
The particle mean free energy 
\be 
\<{m\over 2}{\dot X}^2\> = \int_{-\infty}^{\infty} \rho {m\over 2}(u^2 +v^2)dx=\<K+I\> = {1\over 2}\({\hbar \over 2m\sigma_0}\)^2 
\ee
can be obtained either from direct integration of $\rho {m\over 2}(u^2 +v^2)$, or from noting that the mean work of the stochastic force 
is zero, $\<FX\> = 0$, which for free particle yields $\<m{\ddot X} X\> = 0$. The latter then gives  
\be
\<{m\over 2}{\dot X}^2\> = {m\over 4} \der{^2}{t^2}\<{X}^2\>.
\ee
Since $\<{X}^2\> = \sigma^2$ is the Gaussian variance distribution, (36) provides directly the particle free energy. 

The pressure distribution can be obtained when substituting (34) in (25) and (26)
\be
p =   {\rho \over m}\({\hbar \over 2 \sigma}\)^2 ; \,\,\, 
p_g = {\rho \over m}\({\hbar x \over 2 \sigma^2}\)^2 ; \,\,\,
p_v = {\rho \over m}\({\hbar \over 2 \sigma}\)^2\[ 1- \({x\over \sigma}\)^2\].
\ee
Hence, the internal density energy, $I= (p_g /2 \rho) \propto x^2$, where the vacuum pressure is positive (negative) for values of $|x|$ which are below (above) the standard deviation $\sigma$. To obtain the force partition we substitute (34) in (19)
\be
m{Du\over Dt} = -\der{Q}{x} = m\({\hbar \over 2m}\)^2 {x \over \sigma^4} = -{1\over \rho}\(\der{p_g}{x} + \der{p_v}{x}\) ,
\ee
so that 
\be
-{1\over \rho}\der{p_g}{x} = -{1\over \rho}\der{}{x}\(\rho v^2\) = {Du\over Dt}\[\({x \over \sigma}\)^2 -2\], 
\ee
and
\be
-{1\over \rho}\der{p_v}{x} = {\overline{F} \over m} =  {Du\over Dt}\[3 - \({x \over \sigma}\)^2\].
\ee
Hence, when $|x| < \sqrt{2} \sigma$ the mean particle acceleration is larger than the acceleration of the mean particle velocity 
(i.e. $\overline{F}/m  >  {Du\over Dt}$) and vice-versa for $|x| > \sqrt{2} \sigma$. These terms become equal when $\der{p_g}{x} = 0$ for $\rho = 1/( \sqrt{2\pi}\sigma e)$.

Finally we note that for the dynamic Gaussian solution of (34), equality (32) can be related to the Heisenberg inequality.
This can be shown when recalling that for a free particle (32) gives
\be
 \<(X\dot X)^2 \>  = {\hbar^2\over 2m^2}  -{1\over 3}{d\over dt}{\<X^3 {\dot X} \>}.
\ee
Using then the Cauchy-Schwarz inequality $\sqrt{\<{\dot X}^4 \> \< X^4\>} \ge \<(X\dot X)^2 \>  $, and noting that for Gaussian processes 
$\overline{{d\over dt}{\<X^3 {\dot X} \>}}^t  =0$ (where the time averaging operator $\overline{f(t)}^t \equiv {1\over 2T}\int_{-T}^{T} f(t)dt$ and 
$T \rightarrow \infty$), as well as    
$\<{\dot X}^4 \> = 2\<{\dot X}^2 \>^2$
and $\<{X}^4 \> = 2\<{X}^2 \>^2$, we obtain 
\be
\overline{\sqrt{\<{\dot X}^2 \>}\sqrt{\<{X}^2 \>}}^t \ge {\hbar \over 2m},
\ee
which is inline with the more general results of \cite{Tsekov16} for the stochastic dynamics of a particle in a vacuum.

\section{Discussion}

We find it intriguing that the Schr\"{o}dinger equation can be regarded as an ensemble mean equation of particles that are subject to stochastic forcing. It is
therefore interesting to quantify the mean properties of such stochastic quantum force in hope that these properties will shed light on the fundamental stochastic
law that governs the motion of a particle in a vacuum. The essential result in this paper is that although the expectation value of this force is zero, its local mean does not vanish and is found to be propositional to the third spatial derivative of the probability density function of the particle location. 
The force is uncorrelated with the particle location and hence does not perform work on average. It is however anti-correlated with the volume and this anti-correlation is equal exactly to ${3\hbar^2\over 2m}$, suggesting an Heisenberg-like relation between the volume and the local mean stochastic force.

Furthermore, The stochastic quantum force can be related to a vacuum pressure that is proportional to the PDF second spatial derivative. The latter is accompanied by  the single particle diluted gas pressure which is proportional to the  internal energy and the gas temperature. This pressure partition suggesting that the Bohm potential can be regarded as the the enthalpy associated with the vacuum fluctuations which is equal to the Gibbs energy, as well as to the gas chemical potential. Therefore, the quantum Bohm potential can be interpreted as the energy required to put a particle in a bath of fluctuating vacuum at constant entropy and volume.

The analysis presented here has been performed for the 1D case since the analysis
is relatively straightforward and contains the essence of the physics involved.
The generalization to 3D dynamics is somewhat cumbersome and therefore it was
decided not to be presented in this short paper.\\

%{\noindent \bf Acknowledgments}\\ 
%E.H. is grateful for illuminating discussions with Eliahu Cohen, Massimo Tessarotto,
%Claudio Cremaschini, Michael Berry, Shay Zucker, Dina Prialnik and Attay
%Kovetz.}

\end{document}